\documentclass[12pt,a4paper]{article}
\usepackage{amsfonts,amssymb,amsmath,amsthm,graphicx}

\newcommand{\grad}{\mathop{\rm grad}\nolimits}
\newcommand{\Div}{\mathop{\rm div}\nolimits}

\newtheorem{teo}{Theorem}

\title{Non-stationary parabolic equations for the quasi-monochromatic sound  propagation in media with
a non-stationary background flow}
\author{M.~Yu.~Trofimov}
\date{Il'ichev Pacific oceanological institute FEB RAS\\
Baltiyskaya St. 43, Vladivostok, 690041, Russia\\
e-mail: trofimov@poi.dvo.ru}

\begin{document}

\maketitle

\begin{abstract}
The narrow and wide-angle parabolic equations for the
quasi-monochromatic sound wave packets propagating in a waveguide with a non-stationary
background flow are obtained. The results of numerical simulations are presented.
\end{abstract}

\section{Introduction}
Parabolic equations for sound propagation in moving media were reported by many authors, but
any systematic derivation of such equations on the base of some universal asymptotic method
was not presented in the literature  yet. In this paper we use the generalized multiple-scale
method \cite{nay} to obtain some equations of this type. The material of this paper was presented at the
seminar ``Acoustics of inhomogeneous media X'', Novosibirsk, Russia, 1-6 June 2009.

\section{Derivation of equations}
We start with the equation \cite{ost}
\begin{equation}\label{nonstat}
D_t\frac{\partial}{\partial t}\left(\frac{1}{\rho c^2} D_t p\right) -
\Div\frac{\partial}{\partial t}\left(\frac{1}{\rho}\grad p\right) +
2\cdot\sum_{ij} \frac{\partial \mathbf{V}_{i}}{\partial x_j}\frac{\partial}{\partial x_i}
 \left( \frac{1}{\rho} \frac{\partial p}{\partial x_j}\right)
 = 0 \,,
 \end{equation}
where $\mathbf{V}=(V_1,V_2,V_3)=(u,w,v)$ is the velocity vector of the background flow written on
the orthogonal coordinate system $(x,y,z)$ ($z$-axis is directed upward),
 $D_t=(\partial/\partial t + u\partial/\partial x + w\partial/\partial y +v\partial/\partial z)$,
 $p$ is the acoustic pressure.
\par
Let $\epsilon$ be a small parameter, which is the ratio of characteristic wave length to the characteristic
size of the medium inhomogeneties. Following the generalized multiple-scale metod \cite{nay} we rewrite (\ref{nonstat})
using the slow variables $T=\epsilon t$, $X = \epsilon x$, $Y = \epsilon^{1/2} y$,
$Z = \epsilon^{1/2}z$ and the fast variable $\eta=(1/\epsilon)\theta(X,Y,Z,T)$. We will assume that
the flow velocity is $O(\epsilon)$ and then $u=\epsilon U(X,Y,Z,T)$,
$w=\epsilon^{3/2} W(X,Y,Z,T)$  and $v=\epsilon^{3/2} V(X,Y,Z,T)$, where the quantities $U$, $W$ and $V$ are$O(1)$.
We introduce for convenience the quantity $n=1/c$ and postulate the expansions
\begin{eqnarray*}
n^2 & = & n_0^2(X,Y,T) + \epsilon \nu(X,Y,Z,T)\,,\\
p & = & p_0(X,Y,T,Z,\eta) +   \epsilon p_1(X,Y,T,Z,\eta)+\ldots\,,
\end{eqnarray*}
The density will be assumed depending on the slow variables only, $\rho=\rho(X,Y,Z,T)$. At last, the partial derivatives
are transformed by the rules
$$
\frac{\partial }{\partial x} \rightarrow \epsilon\frac{\partial }{\partial
X} + \theta_X\frac{\partial }{\partial \eta}\,,
\qquad
\frac{\partial }{\partial y} \rightarrow \epsilon^{1/2}\frac{\partial }{\partial
Y} + \epsilon^{-1/2}\theta_Y\frac{\partial }{\partial \eta}\,,
$$
analogously for the variables $t$ è $z$. In the rewritten in such a way (\ref{nonstat}) we collect terms
with like degrees of  $\epsilon$. The squares of the background flow velocities will be neglected as they
was neglected in the derivation of (\ref{nonstat}).
\par
At $O(\epsilon^{-1}$ we obtain that the phase function $\theta$ does not depend on variables $Z$ è $Y$,
$\theta=\theta(X,T)$.
\par
At $O(\epsilon^{0}$ we obtain the Hamilton-Jacobi equation
\begin{equation}\label{h-j}
\left(\theta_X\right)^2 - n_0^2 \left(\theta_T\right)^2  =
\left(\theta_X + n_0 \theta_T\right) \left(\theta_X - n_0
\theta_T\right) =  0\,.
\end{equation}
We shall consider the waves propagating in the positive $X$-direction and adopt as the Hamilton-Jacoby equation the first
factor in (\ref{h-j}).
\par
The equation at $O(\epsilon^{1})$ does not contain $p_1$ on the strength of (\ref{h-j}).
Substituting in this equation the anzats $p_0 =A_0(X,Y,T,Z)\exp(\mathrm{i}\eta)$, we obtain the equation for
the amplitude $A_0$
\begin{equation}\label{parA}
2\mathrm{i}n_0\omega\frac{1}{\rho}\left[A_{0,X} + n_0 A_{0,T}\right] +
\left(\frac{1}{\rho}A_{0,Y}\right)_Y +
\left(\frac{1}{\rho}A_{0,Z}\right)_Z + \chi A_0 = 0\,,
\end{equation}
where $\omega=-\theta_T$ is the local frequency of sound,
\begin{equation}\label{coef}
\chi = \left\{\left[\left(\frac{1}{\rho}n_0 \right)_X +
n_0\left(\frac{1}{\rho}n_0 \right)_T\right]\omega \mathrm{i} +
\frac{1}{\rho}\left(\nu-2 n_0^3 U\right) \omega^2\right\}\,.
\end{equation}
Assuming that
$$
\nu=-2\frac{c_1}{c_0^3}\,,
$$
where $c_0=1/n_0$, we obtain that
$$
\frac{1}{\rho}\omega^2\nu-2\frac{1}{\rho}\omega^2 n_0^3 U=2\frac{1}{\rho}\omega^2\frac{c_1+U}{c_0^3}\,,
$$
so  the potential of the parabolic equation contains the so called {\em effective sound speed}
$c+U$,
as was first shown in the paper \cite{nghiem}. We see that our equation is a generalization of the parabolic equation
obtained in that work.
\par
Note that the dependence on variable $Y$ in the obtained equation is just the same as on $Z$.
Therefore in the sequel  we shall not write the terms expressing the dependence on
 $Y$ and consider the $2D$ waveguide. As will be easily seen, all result can be directly transferred to the
 $3D$ case.
\par
The analogous considerations at $O(\epsilon^{2})$ gives the equation
for the amplitude $A_1$, $p_1 = A_1(X,Z,T)\exp(\mathrm{i}\eta)$.
\begin{equation}\label{parB}
2\mathrm{i}n_0\omega\frac{1}{\rho}\left[A_{1X} + n_0 A_{1T}\right] +
\left(\frac{1}{\rho}A_{1Z}\right)_Z + \chi A_1 + {\cal F} = 0\,,
\end{equation}
where
\begin{equation}\label{parBF}
\begin{split}
{\cal F} =& \left(\frac{1}{\rho}A_{0X}\right)_X -
\left(\frac{1}{\rho}n_0^2 A_{0T}\right)_T + 2\mathrm{i}\frac{1}{\rho}\nu\omega A_{0T}
+ \mathrm{i}\left(\frac{1}{\rho}\nu\omega_T\right)_T A_{0}\\
&- \frac{\mathrm{i}}{\omega}U\left(\frac{1}{\rho}n_0^2\omega^2 A_0\right)_X -3\mathrm{i}U\frac{1}{\rho}n_0^2\omega
 A_{0X} - 2\mathrm{i}U_X\frac{1}{\rho}n_0^2\omega A_0\\
 &+ \mathrm{i}\left[\left(\frac{1}{\rho}n_0\right)_X- \frac{1}{\rho}n_0 n_{0T}\right]Un_0\omega A_0 -
\mathrm{i}\frac{1}{\omega}V\left(n_0^2\omega^2\frac{1}{\rho}A_0\right)_Z\\
&- \mathrm{i}\frac{1}{\rho}n_0^2\omega VA_{0Z} - Un_0\left(\frac{1}{\rho} A_{0Z}\right)_Z -
2U_Z \frac{1}{\rho} A_{0Z}  \,,
 \end{split}
\end{equation}
It can be shown \cite{tr} that the system of equations (\ref{parA}),
(\ref{parB}) generalize the known wide-angle stationary parabolic
equation obtained by the factorization method with the
rational-linear Pad\'e approximation of the square root operator.
Moreover, even in the stationary case our equations contains the
terms that the factorization method cannot produce.

\section{Initial-boundary value problems for the pa\-ra\-bo\-lic equations}\sloppy
For the simulation of the sound waves in the ocean the most interesting initial-boundary value problems
for the Hamilton-Jacobi equatuion (\ref{h-j}) and the parabolic equations
  (\ref{parA}), (\ref{parB}) are the problems
with  $X$ as the evolution variable in the domain
$\Omega = \{Z_0<Z<Z_1\}\times \{T_0<T<T_1\}$.
\par
For the energy norm
\begin{equation*}\label{norm}
E(X) =  \int_{T_0}^{T_1}\int_{Z_0}^{Z_1} \frac{1}{\rho} |\theta_X A_0|^2\,dZdT\,.
\end{equation*}
under some simple and natural assumptions on the boundary conditions
at
$\partial\Omega$, the following theorem holds
\begin{teo}\label{te2}
The energy norm of the solution of the initial-boundary value problem for the parabolic equation (\ref{parA})
satisfies the inequality
\begin{equation}\label{enerIn}
E(X) \le E(0)\cdot\exp\left(\int_{0}^{X}\sup_T \left(\frac{n_{0,s}(s,T)}{n_0(s,T)} \right)\,ds\right)
%\int_{T_0}^{T_1}\frac{n_{0,X}}{n_0}\int_{Z_0}^{Z_1}\frac{1}{\rho} |\theta_X A_0|^2\,dZdT\,.
\end{equation}
If $n_0$ can be represented in the form $n_0=\bar n_0(X) a(T)$ then the following inequality holds
\begin{equation}\label{enerIn1}
E(X) \le E(0)\frac{\bar n_0(X)}{\bar n_0(0)}\sup_T a(T)\,.
\end{equation}
\end{teo}
This theorem immediately implies the uniqueness of solutions of such initial-boundary value problems for
(\ref{parA}), (\ref{parB}) in functional spaces of the type $C([0,X],L_2(\Omega))$.

\section{Numerical simulation}
\begin{figure}[h]
\begin{center}
\includegraphics[width=0.49\textwidth]{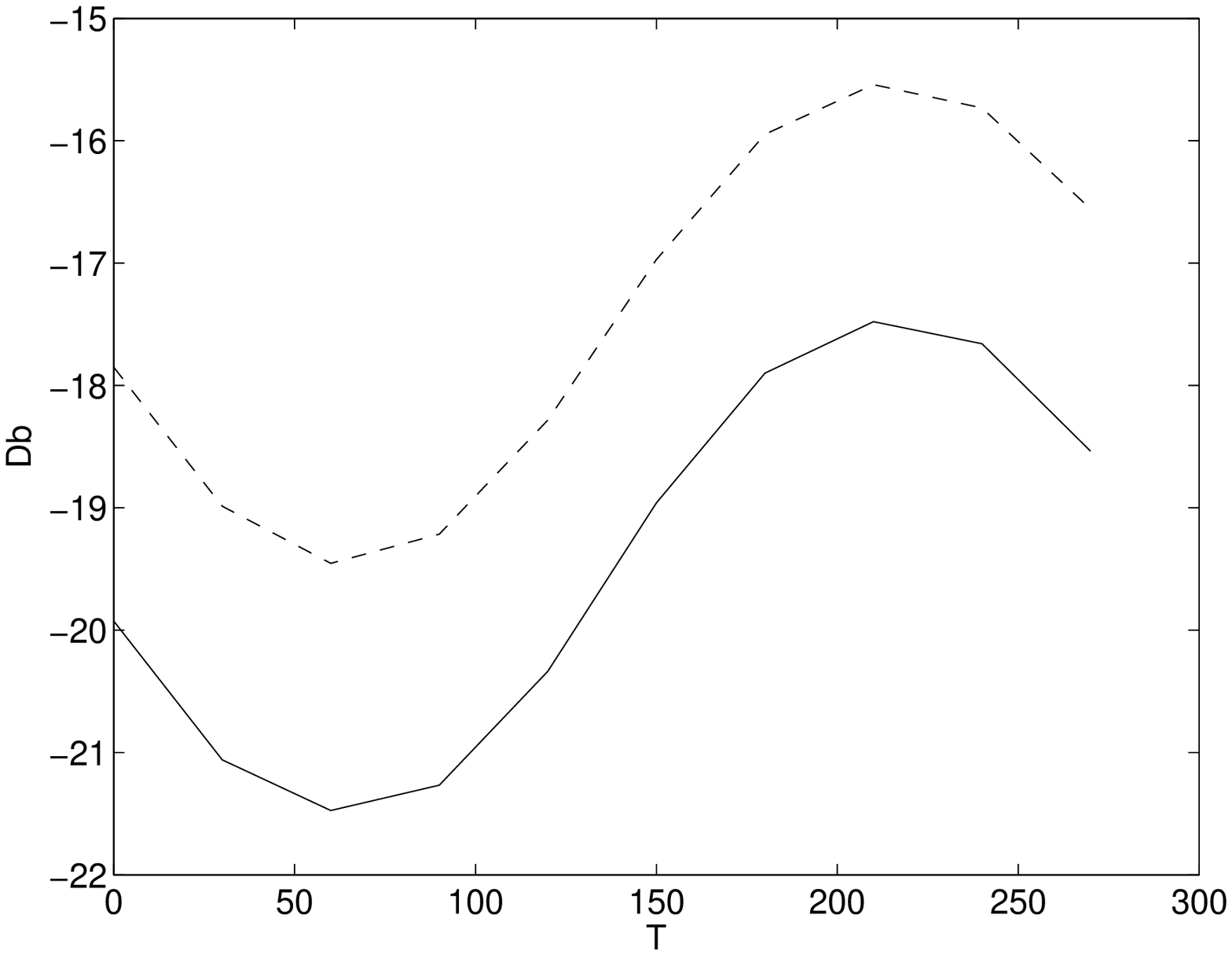}
\includegraphics[width=0.5\textwidth]{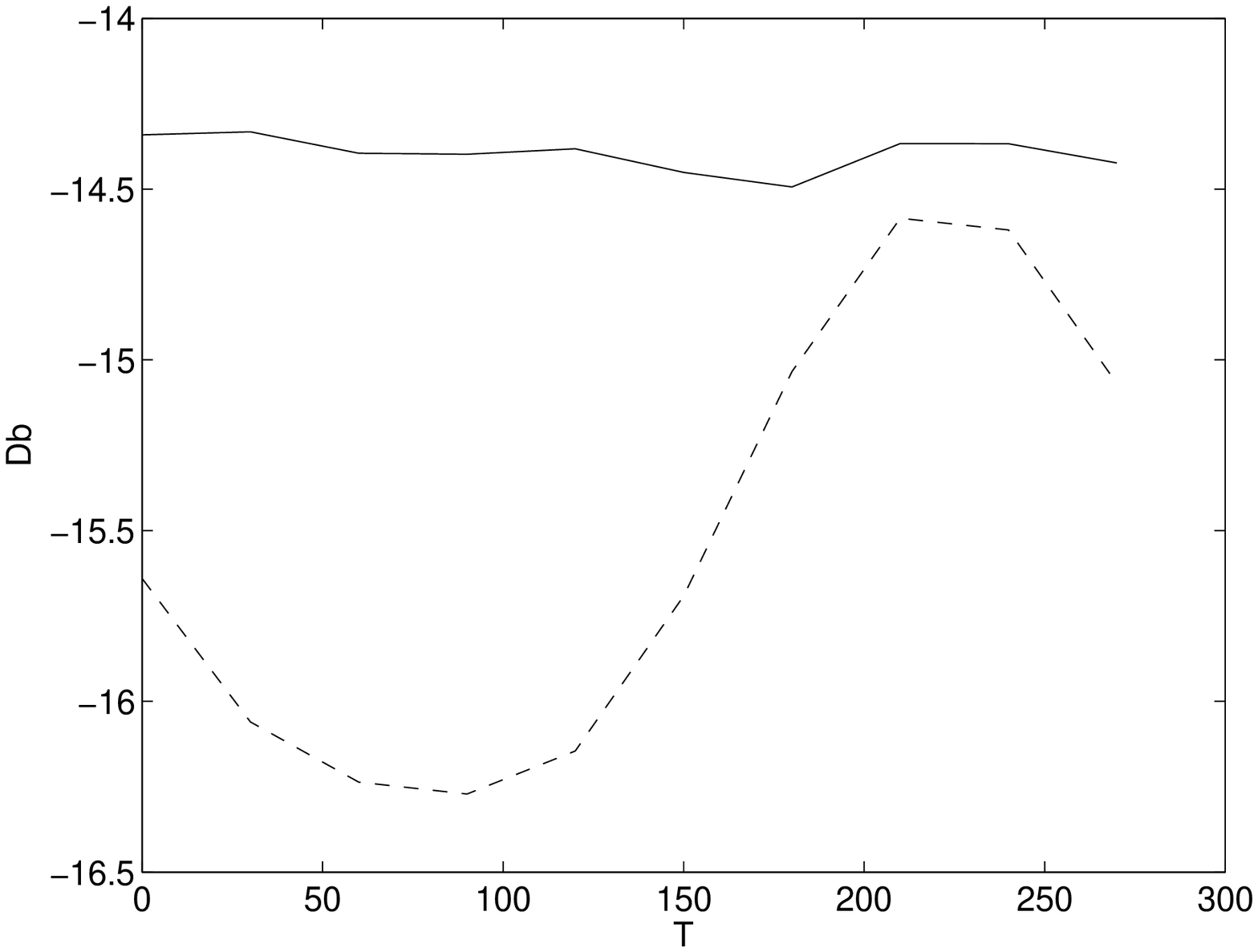}
\caption{Temporal variations of the transmission loss at the distance 6~km (left) and 12~km (right):
with the velocity terms (dashed line), without these terms (solid line). The second mode of internal waves.}
\label{npa36}
\end{center}
\end{figure}

\begin{figure}[h]
\begin{center}
\includegraphics[width=0.5\textwidth]{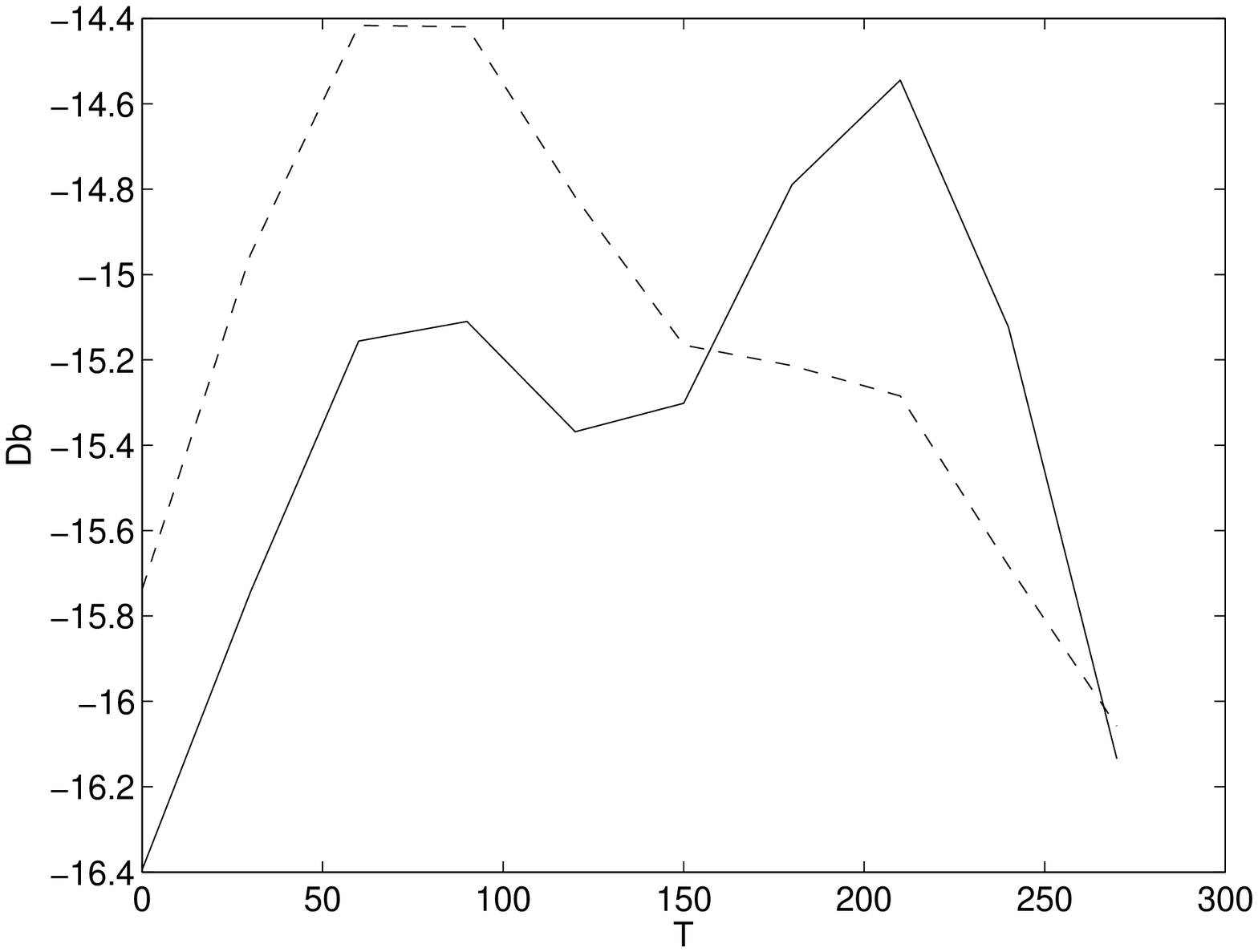}
\includegraphics[width=0.49\textwidth]{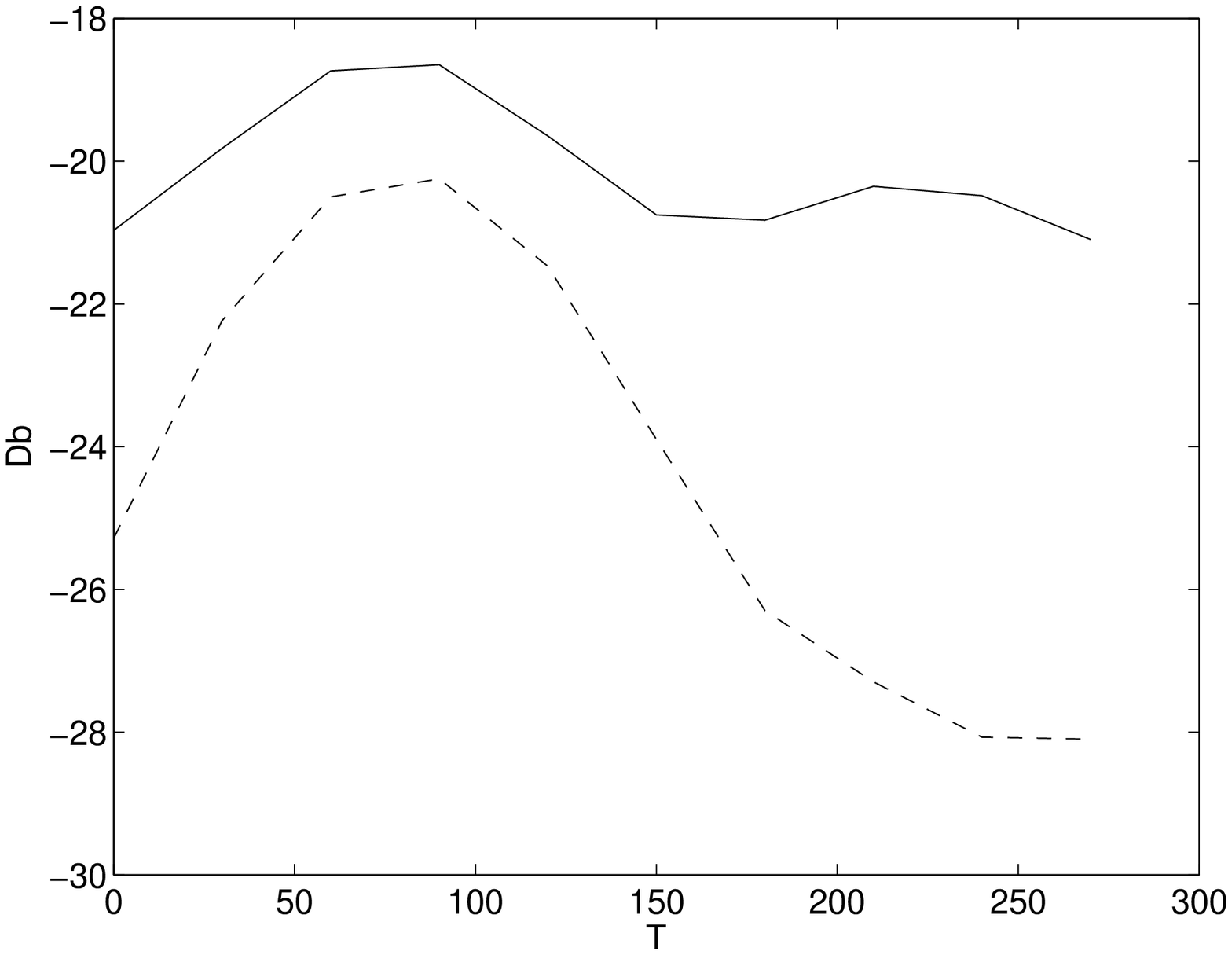}
\caption{Temporal variations of the transmission loss at the distance 6~km (left) and 12~km (right):
with the velocity terms (dashed line), without these terms (solid line). The third mode of internal waves.}
\label{npa26}
\end{center}
\end{figure}

In the parabolic equations (\ref{parA}), (\ref{parB}) the influence of the background flow is taken into account
not only through the terms which explicitly contain the corresponding velocities, but also through the deformation
of the sound speed and density profiles. As the geometric small parameter $\epsilon$ in typical cases is much
larger than the Mach number for the ocean medium, the question about essentiality of introducing the background flow
velocities in the narrow-angle parabolic equation (\ref{parA}) is naturally risen.

\par
For an answer to this question the simulation of the harmonic sound wave propagation through the harmonic internal
wave of a given mode was conducted, with the following parameters: the undisturbed waveguide of constant
 depth (100~m) was given by the parabolic sound speed profile with the minimum (1460~m/s) at the center of the
 waveguide and maximum (1500~m/s) at the top and bottom boundaries, the sound frequency was taken to be equal 100~Hz
 and the boundary conditions for the sound field was taken to be soft at the top boundary and hard at the bottom boundary.
 The point sound source was situated at the center of the waveguide.
 The density stratification was given by the constant Brent-V\"as\"al\"a frequency ($1/127~\text{s}^{-1}$), for the
 simulation  were used the five minutes period internal waves of the second mode (wavelength=214~m) and the third mode
 (wavelength=142~m).

\par
The results are presented in fig.~\ref{npa36},\ref{npa26} in the form of the temporal variations of the transmission loss
at the distances of 6~km and 12~km from the source. We can conclude that the explicit introduction of the background
flow velocities to the narrow-angle parabolic equation is essential.

\section{Conclusion}
In this paper the system of parabolic-like equations (\ref{parA}), (\ref{parB}), which can be used for
numerical modelling of sound propagation in waveguides with non-stationary background flow, is obtained.
Some properties of these equations are established. We hope that this information will be useful for the computational
acoustics community. Complete derivations, proofs and more extensive numerical modelling will be presented in the
forthcoming paper.

\end{document}